\documentclass[twocolumn,english,aps,prb,showpacs,superscriptaddress]{revtex4}
\usepackage[T1]{fontenc}
\usepackage{amsmath,graphicx,amssymb,dsfont,color}

\def\bbm[#1]{\mbox{\boldmath $#1$}}
\renewcommand{\Re}{\text{Re}}
\renewcommand{\Im}{\text{Im}}

\begin{document}

\title{Tuning the electromagnetic local density of states in graphene-covered systems\\via strong coupling with graphene plasmons}

\author{Riccardo Messina}\email{riccardo.messina@institutoptique.fr}
\author{Jean-Paul Hugonin}
\author{Jean-Jacques Greffet}
\author{Fran\c{c}ois Marquier}
\affiliation{Laboratoire Charles Fabry, UMR 8501, Institut d'Optique, CNRS, Universit\'{e} Paris-Sud 11,\\2, Avenue Augustin Fresnel, 91127 Palaiseau Cedex, France.}
\author{Yannick De Wilde}
\affiliation{Institut Langevin, ESPCI ParisTech, CNRS, 1 rue Jussieu, 75005 Paris, France.}
\author{Ali Belarouci}
\author{Luc Frechette}
\affiliation{Laboratoire Nanotechnologies Nanosyst\`{e}mes (LN2), UMI CNRS 3463, Universit\'{e} de Sherbrook, Canada.}
\author{Yvon Cordier}
\affiliation{Centre de Recherche sur l'H\'{e}t\'{e}ro-Epitaxie et ses Applications, Centre National de la Recherche Scientifique, Rue Bernard Gregory, Sophia Antipolis, F-06560 Valbonne, France.}
\author{Philippe Ben-Abdallah}\email{pba@institutoptique.fr}
\affiliation{Laboratoire Charles Fabry, UMR 8501, Institut d'Optique, CNRS, Universit\'{e} Paris-Sud 11,\\2, Avenue Augustin Fresnel, 91127 Palaiseau Cedex, France.}

\date{\today}

\pacs{73.20.Mf, 07.79.Fc, 44.40.+a, 78.20.-e}

\begin{abstract}
It is known that the near-field spectrum of the local density of states of the electromagnetic field above a SiC/air interface displays an intense narrow peak due to the presence of a surface polariton. It has been recently shown that this surface wave can be strongly coupled with the sheet plasmon of graphene in graphene-SiC heterosystems. Here, we explore the interplay between these two phenomena and demonstrate that the spectrum of the electromagnetic local density of states in these systems presents two peaks whose position depends dramatically both on the distance to the interface and on the chemical potential of graphene. This paves the way towards the active control of the local density of states.
\end{abstract}

\maketitle

\section{Introduction}

\begin{figure}[h!]
\scalebox{0.12}{\includegraphics{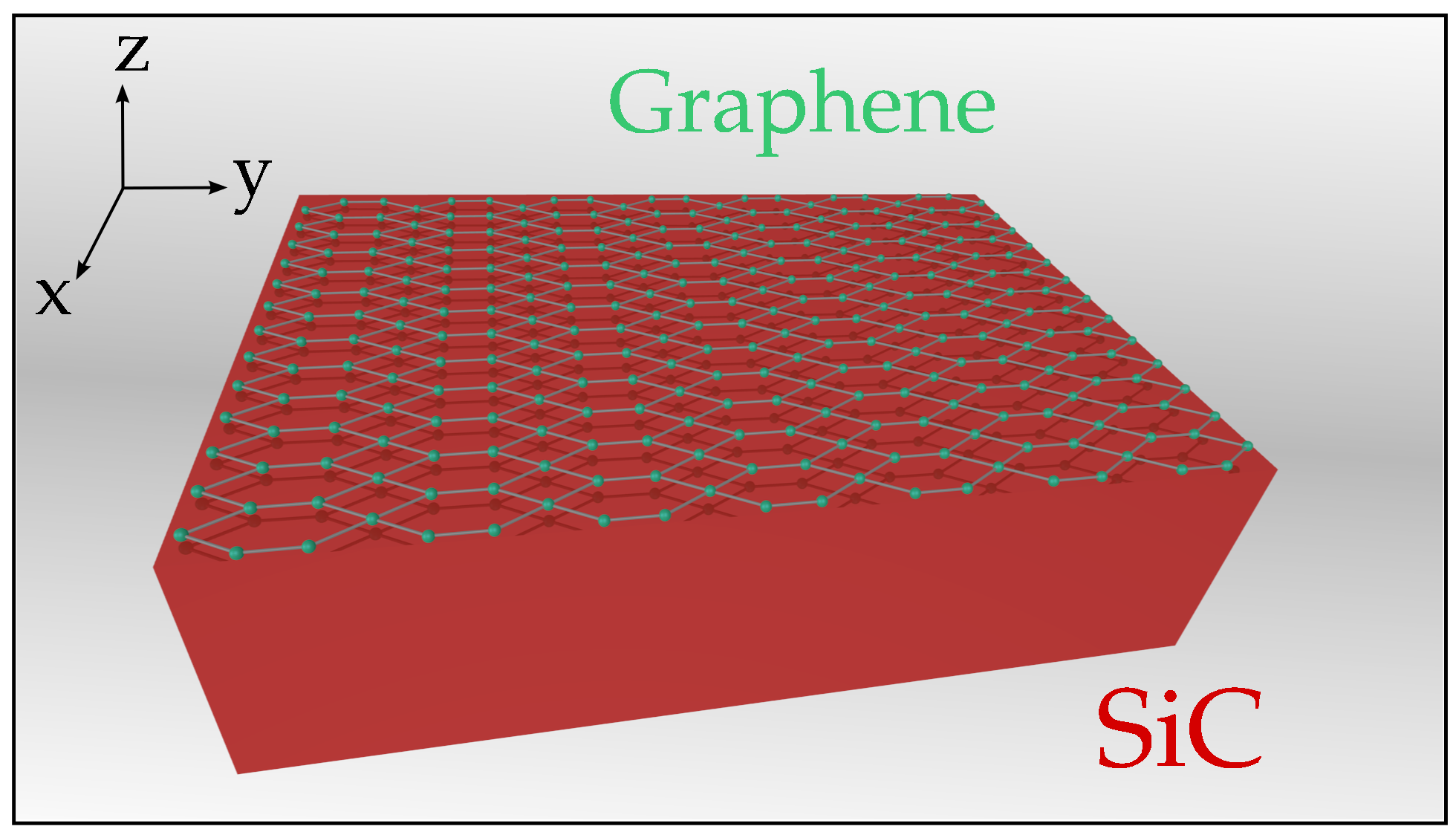}}
\caption{Scheme of the system. A dielectric (SiC) substrate is covered with a monolayer graphene sheet coinciding with the plane $z=0$.}
\label{FigSys}\end{figure}

Graphene has attracted during the last years a strong interest because of its unique electronic \cite{Geim1,Geim2,Avouris,BatzillSurfSciRep12,AndreiRepProgPhys12} and optical properties \cite{Geim3} that could lead to breakthrough technologies in the domains of nanoelectronics and nano-optics. One of the main features is the linear form of graphene electronic dispersion curves near the edges of the Brillouin zone, which is the signature of charge carriers of zero effective mass. More recently, the study of graphene optical properties has provoked a very particular attention \cite{MikhailovPRL07,KoppensNanoLett11,VakilScience11,NikitinPRB11,ThongrattanasiriPRL12}. An important property of monolayer graphene is the presence of strongly confined sheet plasmons \cite{LangerNewJPhys11,PolitanoPRB12} which can be easily tuned by modifying the charge density within the lattice of carbon atoms \cite{PerssonJPhysCondensMatter10,VolokitinPRB11,FalkovskyJPhysConfSer08,JablanPRB09,KochPRB10}. It has been suggested that these tunable surface modes could be used to manage near-field radiative heat transfer \cite{SvetovoyPRB12,IlicPRB12} or to convert near-field energy with graphene-based thermophotovoltaic devices \cite{IlicOptExpress12,MessinaArxiv12}. When studying these surface excitations, the intrinsic quantity containing all the relevant information is the electromagnetic local density of states (EM-LDOS) \cite{JoulainPRB03}. Let us remind the meaning of the electromagnetic density of states in vacuum: it is used to describe the spontaneous emission rate as well as the blackbody energy density. In the presence of interfaces both quantities are modified. This is due to both the presence of surface modes and to interferences between incident and reflected waves. These processes result in a space-dependent density of states. This is conveniently described by means of the EM-LDOS. Surface modes have been recently shown to affect significantly the local equilibrium energy density \cite{ShchegrovPRL00,DeWildeNature06,JonesNanoLett12} and also to modify spontaneous emission rate \cite{DrexhageProgOpt74,AngerPRL06,KuhnPRL06}. The EM-LDOS also plays a key role in the study of radiative cooling of a nanoparticle in proximity of a surface \cite{TschikinEurPhysJB12}, losses in atomic traps \cite{HenkelEurophysLett99,HarberJLowTempPhys03}, as well as to derive the Casimir force \cite{VanKampenPhysLettA68}. Tuning the EM-LDOS at the surface of nanostructured materials is also a challenging problem \cite{BenAbdallahApplPhysLett09}. Here we focus on the particular configuration of a multilayered planar structure. In this case, the EM-LDOS can be decomposed in an electric and a magnetic contribution as $\rho=\rho_\text{E}+\rho_\text{M}$, where \cite{JoulainPRB03}
\begin{eqnarray}\label{EM-LDOSe}\nonumber&&\hspace{-.3cm}\rho_\text{E}(d,\omega)=\frac{\rho_v(\omega)}{4}\Bigl\{2+\int_0^1\frac{\kappa\,d\kappa}{p}\Bigl[\Re\bigl(r_\text{TE}e^{2i\frac{\omega}{c}pd}\bigr)\\
&+&(2\kappa^2-1)\Re\bigl(r_\text{TM}e^{2i\frac{\omega}{c}pd}\bigr)\Bigr]\\
\nonumber&+&\int_1^{+\infty}\frac{\kappa\,d\kappa}{|p|}e^{-2\frac{\omega}{c}|p|d}\Bigl[\Im\bigl(r_\text{TE}\bigr)+\Im\bigl(r_\text{TM}\bigr)(2\kappa^2-1)\Bigr]\Bigr\},\end{eqnarray}
and $\rho_\text{M}$ is obtained from $\rho_\text{E}$ by exchanging the polarizations TE and TM. In this expression $d$ is the distance from the surface along the $z$ axis (the surface coincides with the $xy$ plane, see Fig. \ref{FigSys}), $\kappa=ck/\omega$, $k$ being the component of the wavevector parallel to the surface, $p=\sqrt{1-\kappa^2}$ and $\rho_v(\omega)=\omega^2/\pi^2c^3$ is the EM-LDOS in vacuum. The material properties of the surface are taken into account in the EM-LDOS \eqref{EM-LDOSe} through the Fresnel reflection coefficients $r_\text{TE}$ and $r_\text{TM}$ for the two polarizations.

Several experimental techniques can be used to probe the near-field properties of a surface. The dispersion relation of sheet plasmons can be studied using electron energy loss spectroscopy (EELS) \cite{Ibach82,LiuPRB10,LangerNewJPhys12,PolitanoPRB11}. This technique, used in conjunction with a scanning tunneling electron microscope (STEM), has recently allowed a measurement of the EM-LDOS \cite{NelayahNatPhys07}. As analyzed in ref. \cite{GarciaDeAbajoPRL08}, STEM-EELS yields a signal related to the integral of the EM-LDOS along the trajectory of the electrons. As a consequence, the $z$ dependence of the EM-LDOS cannot be observed using STEM-EELS. Direct observation of sheet plasmons on graphene has also been reported recently \cite{FeiNanoLett11,ChenNature12,FeiNature12} using a scanning near-field optical microscope (SNOM). This kind of technique (the analogue of scanning tunneling microscope for electronic LDOS) consists in approaching an AFM tip to the surface, which scatters the evanescent field in proximity of the surface into a propagative contribution, observed by a far-field detector. This conversion is usually theoretically described by representing the tip as a point dipole \cite{JoulainPRB03}. By allowing the control of the distance between the tip and the surface, this technique allows the study of the $z$ dependence of the EM-LDOS. The connection between the observed signal and the EM-LDOS depends indeed on the details of the experimental techniques, as discussed in \cite{FeiNanoLett11,ChenNature12,JoulainPRB03,GarciaDeAbajoPRL08,DeWildeNature06,JonesNanoLett12}.

In this paper, our attention is focused on the distance dependence of the EM-LDOS. It has been shown \cite{ShchegrovPRL00} that the EM-LDOS has dramatic spectral changes in the near field due to to the resonant contribution of the surface waves at the frequencies where $\epsilon(\omega)+1=0$.  These large spectral changes with distance in vacuum are a spectacular example of the non-invariance of spectra upon propagation in vacuum first discussed in ref. \cite{WolfPRL86}. We consider here a graphene sheet deposited on silicon carbide (SiC) (see Fig. \ref{FigSys}) and show that the EM-LDOS peak positions strongly depend on distance. In addition, they also depend on chemical potential, paving the way to an active control of the EM-LDOS. In particular we pay a specific attention to the description of the coupling mechanism between the graphene plasmon and the surface phonon-polariton supported by the SiC and highlight the fingerprint of this coupling on the EM-LDOS spectrum. This paper is organized as follows. In Section \ref{SecEMLDOS} we discuss the behavior of the electromagnetic local density of states for SiC alone, for a suspended graphene sheet, and for the coupled system SiC-graphene. Then, in Section \ref{SecDispRel} these results are interpreted in terms of the dispersion relations for the surface modes in the three different configurations. Finally, Section \ref{SecConcl} contains some conclusive remarks.

\section{Electromagnetic local density\newline of states}\label{SecEMLDOS}

In the general configuration of a substrate having a frequency-dependent dielectric permittivity $\varepsilon(\omega)$ covered with graphene (having 2D surface conductivity $\sigma(\omega)$) the reflection coefficients appearing in eq. \eqref{EM-LDOSe} take the form \cite{FalkovskyJPhysConfSer08,Geim3}
\begin{equation}\label{Fresnel}\begin{split}r_\text{TE}&=\frac{p-p_m-\mu_0c\,\sigma(\omega)}{p+p_m+\mu_0c\,\sigma(\omega)},\\
r_\text{TM}&=\frac{\varepsilon(\omega)p-p_m+\mu_0c\,\sigma(\omega)p\,p_m}{\varepsilon(\omega)p+p_m+\mu_0c\,\sigma(\omega)p\,p_m},\end{split}\end{equation}
where $p_m=\sqrt{\varepsilon(\omega)-\kappa^2}$ equals $c/\omega$ times the $z$ component of the wavevector inside the medium. The ordinary case of a dielectric substrate is recovered by taking $\sigma(\omega)=0$, whereas $\epsilon(\omega)=1$ gives back the reflection coefficients of a suspended graphene sheet.

Before calculating the EM-LDOS, we need a model for the permittivity $\varepsilon(\omega)$ of SiC and for the conductivity $\sigma(\omega)$ of graphene. For the permittivity we use a Lorentz model \cite{Palik98} $\varepsilon(\omega)=\varepsilon_\infty(\omega^2-\omega_L^2+i\Gamma\omega)/(\omega^2-\omega_T^2+i\Gamma\omega)$ with $\varepsilon_\infty=6.7$, $\omega_L=1.827\times10^{14}\,\text{rad\,s}^{-1}$, $\omega_T=1.495\times10^{14}\,\text{rad\,s}^{-1}$ and $\Gamma=0.9\times10^{12}\,\text{rad\,s}^{-1}$. This model predicts the existence of a surface phonon-polariton resonance at frequency $\omega_\text{spp}\simeq1.787\times10^{14}\,\text{rad s}^{-1}$. As for the conductivity $\sigma(\omega)$, it can be written as a sum of an intraband (Drude) and an interband contribution, respectively given by \cite{FalkovskyJPhysConfSer08}
\begin{eqnarray}\label{Sigma}&&\sigma_D(\omega)=\frac{i}{\omega+\frac{i}{\tau}}\frac{2e^2k_BT}{\pi\hbar^2}\log\Bigl(2\cosh\frac{\mu}{2k_BT}\Bigr),\\
&&\nonumber\sigma_I(\omega)=\frac{e^2}{4\hbar}\Bigl[G\Bigl(\frac{\hbar\omega}{2}\Bigr)+i\frac{4\hbar\omega}{\pi}\int_0^{+\infty}\frac{G(\xi)-G\bigl(\frac{\hbar\omega}{2}\bigr)}{(\hbar\omega)^2-4\xi^2}\,d\xi\Bigr],\end{eqnarray}
where $G(x)=\sinh(x/k_BT)/[\cosh(\mu/k_BT)+\cosh(x/k_BT)]$. The conductivity depends explicitly on the temperature $T$ of the graphene sheet, for which we have chosen $T=300\,$K in our calculations. Moreover, it contains the chemical potential $\mu$, which represents a varying parameter in our discussion, and the relaxation time $\tau$, for which we have used the value \cite{JablanPRB09} $\tau=10^{-13}\,$s.

\begin{figure}[h!]
\scalebox{0.051}{\includegraphics{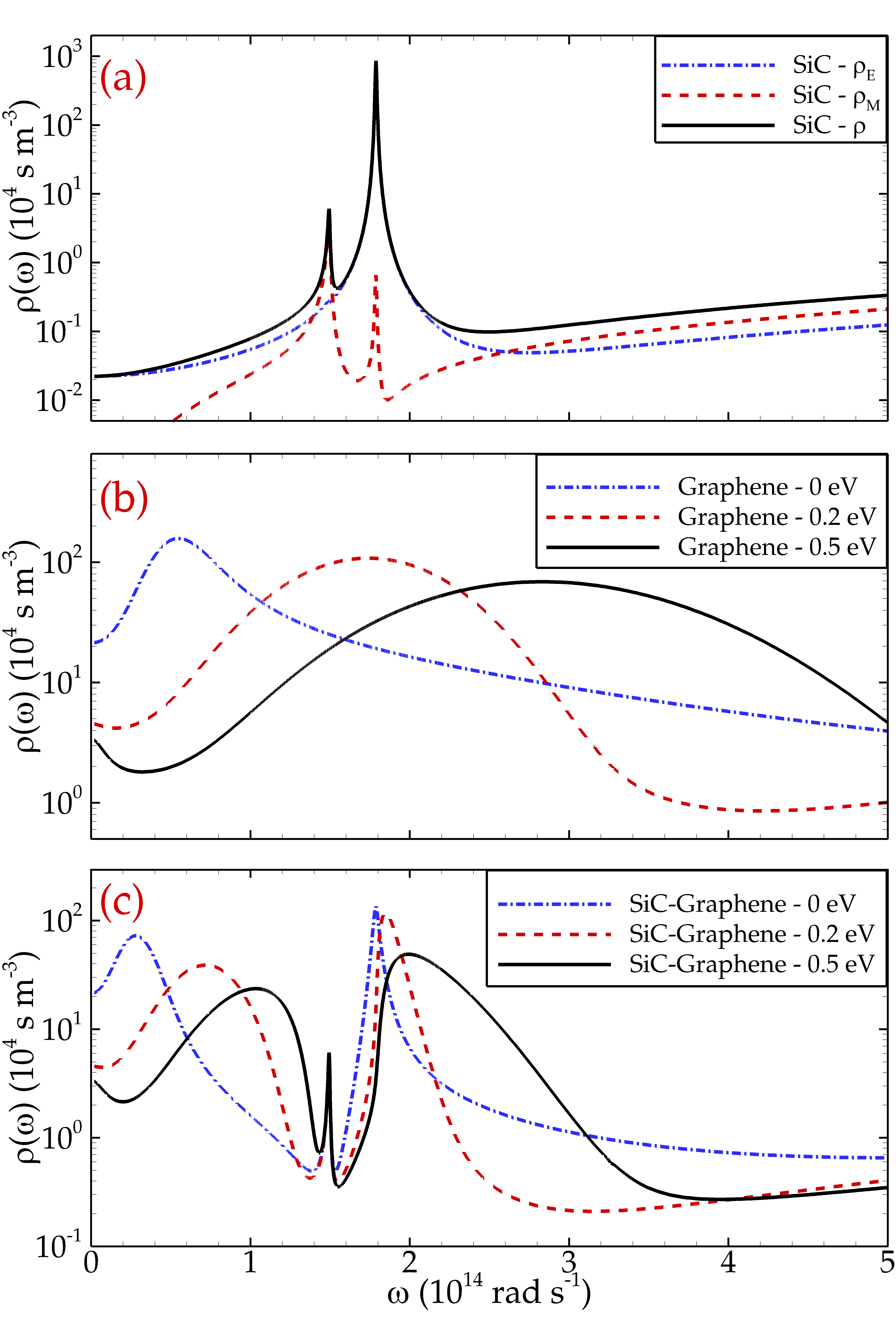}}
\caption{EM-LDOS at a distance of $d=50\,$nm from a SiC substrate (a), a suspended graphene sheet (b) and a graphene-covered SiC substrate (c). In panel (a) the total EM-LDOS (solid line) is decomposed in its electric (dot-dashed line) and magnetic contribution (dotted line). In panels (b) and (c), corresponding to suspended graphene and graphene-covered SiC respectively, only the total EM-LDOS is represented, for three values of the graphene chemical potential: $\mu=0\,$eV (blue dot-dashed line), $\mu=0.2\,$eV (red dashed line) and $\mu=0.5\,$eV (black solid line).}
\label{FigEM-LDOS}\end{figure}

We now turn to the analysis of the EM-LDOS spectra. In Fig. \ref{FigEM-LDOS} we show the electric, magnetic and total EM-LDOS for a semi-infinite SiC surface (Fig. \ref{FigEM-LDOS}(a)), for one layer of suspended graphene (Fig. \ref{FigEM-LDOS}(b)) and for  graphene on SiC (Fig. \ref{FigEM-LDOS}(c)). All the EM-LDOS shown in Fig. \ref{FigEM-LDOS} are calculated at a distance of $d=50\,$nm from the surface, chosen to be smaller than the decay distance in air of SiC surface polaritons.
The curves in Fig. \ref{FigEM-LDOS}(a) show the well-known peak \cite{ShchegrovPRL00} at $\omega=\omega_\text{spp}$ due to the surface phonon polariton contribution, almost entirely associated to the electric contribution. The magnetic contribution, on the contrary, is at the origin of a secondary peak at $\omega=\omega_T$, approximately two orders of magnitude weaker, associated to the resonance of the dielectric constant of SiC \cite{Biehs}. We now focus on a suspended sheet of graphene (Fig. \ref{FigEM-LDOS}(b)). In this case, we have observed that the electric part of the EM-LDOS is manifestly dominating for any considered frequency: for this reason, we only represent the total EM-LDOS both for suspended graphene and graphene-covered SiC. Note that the density of states is very broad as compared to the case of SiC and shows only one maximum less pronounced than for SiC. A remarkable property of graphene is the possibility of tuning the peak frequency by tuning the chemical potential $\mu$. In particular, for the intermediate value $\mu=0.2\,$eV, the peak matches the phonon-polariton SiC resonance. We are now interested in discussing what happens when graphene is deposited on SiC (Fig. \ref{FigEM-LDOS}(c)). For all considered $\mu$, we observe the appearance of three peaks in the EM-LDOS. One of them appears at  $\omega=\omega_T$. Its height and position do not depend on $\mu$. This peak is due to $s$-polarized magnetic fields so that it is not associated with surface phonon polaritons and graphene plasmons. We now focus on the two other peaks. One of them keeps the memory of the SiC surface phonon-polariton resonance but is shifted as $\mu$ increases and becomes broader. The other peak also shifts toward higher frequencies as $\mu$ increases.

\section{Dispersion relation of surface modes and strong coupling}\label{SecDispRel}

We now discuss how in each considered configuration the EM-LDOS results from the surface modes interacting within the graphene film. To this aim we study the poles of the reflection coefficients \eqref{Fresnel}, by focusing in particular on how the optical properties of graphene modify the shape of the surface modes of SiC. This phenomenon will be specifically discussed, as the EM-LDOS presented in Fig. \ref{FigEM-LDOS}, as a function of $\mu$. It is well known \cite{Raether} that the optical description of a polar material by means of a Lorentz model predicts the existence of surface modes only in TM polarization. For this reason we will limit our analysis to TM polarization also in the case of suspended graphene and graphene on SiC. In order to derive the dispersion relation of surface modes in the three cases, we study the zeros of the denominator of $r_\text{TM}$ (see eq. \eqref{Fresnel}). It is known that the dispersion relation depends on the choice made when searching the poles. As discussed in ref. \cite{ArchambaultPRB09}, when discussing EM-LDOS, we look for poles with real wavevector $k$ and complex frequency $\omega=\omega_r+i\omega_i$. In this calculation scheme, $\omega_r$ provides the energy of the considered mode, while the inverse of $\omega_i$ defines its lifetime. We have performed this calculation for the three configurations in the same frequency region in which the EM-LDOS has been discussed, and for $k$ varying between 0 and 100\,$\mu$m$^{-1}$. This choice is associated to the fact that for a given distance $d$ from the surface, the modes participating to the density of states are smaller or of the order of \cite{BenAbdallahPRB10} $k_c\simeq d^{-1}$, which for $d=50\,$nm gives $k_c\simeq 20\,\mu$m$^{-1}$.

\begin{figure}[h!]
\scalebox{0.052}{\includegraphics{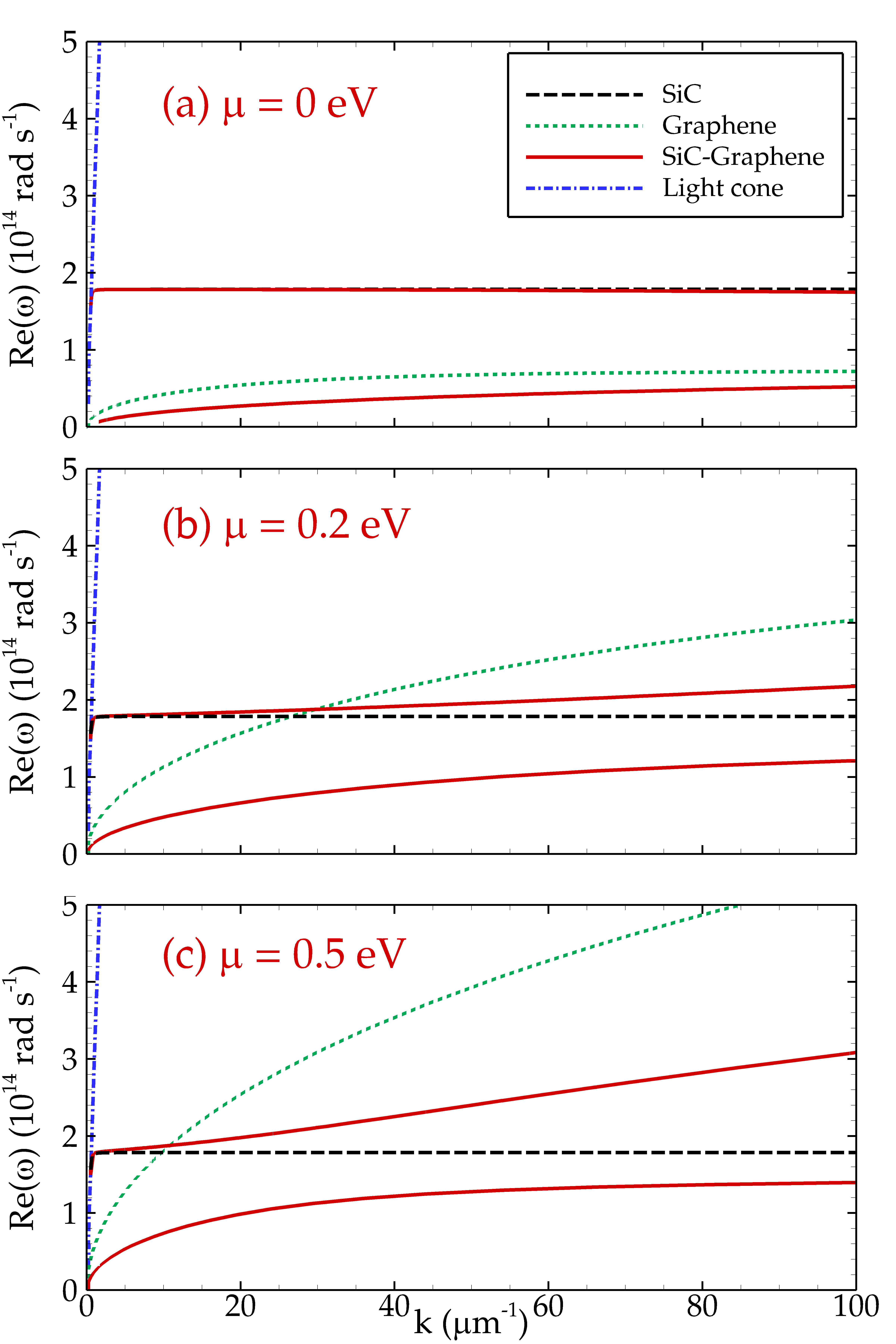}}
\caption{Dispersion relations of surface modes for SiC (black dashed line), suspended graphene (green dotted line) and graphene-covered SiC (red solid line). The blue dot-dashed line represents the light cone in vacuum. The comparison is made for three different values of the graphene chemical potential ($\mu=0\text{(a)},0.2\text{(b)},0.5\text{(c)}\,$eV).}
\label{FigDR}\end{figure}

The results are shown in Fig. \ref{FigDR} for the three configurations and the same three choices of $\mu$ used for the EM-LDOS.
The behavior of a Lorentz material in terms of surface-mode dispersion relation is well-known: it implies a branch following first (for small frequencies) the light cone, then approaching a horizontal asymptote at $\omega=\omega_\text{spp}$ \cite{ShchegrovPRL00}. For SiC, this asymptote is reasonably reached around $k=2\,\mu$m$^{-1}$. Thus, in the wavevector scale of Fig. \ref{FigDR} the SiC curve reduces as a matter of fact to a horizontal (dashed) line at $\omega=\omega_\text{spp}$. This horizontal asymptote is at the origin of the pronounced near-field peak at $\omega=\omega_\text{spp}$ in the EM-LDOS associated to SiC alone \cite{JoulainSurfSciRep05}.

Let us now focus on suspended graphene. In this case, for any $\mu$ we observe the already known characteristic $\sqrt{k}$-like dispersion relation \cite{FalkovskyJPhysConfSer08}. Nevertheless, this analytical dependence of $\omega$ on $k$ is deduced by performing several approximations, and remarkably by taking into account only the intraband contribution $\sigma_D(\omega)$ to the conductivity (see eq. \eqref{Sigma}). This simplification starts failing when reducing the value of $\mu$ and increasing the value of $\omega$, even in the domains we are considering in this work. The result is a dispersion relation which has a similar shape, but increases always more slowly than the curve obtained for $\sigma(\omega)=\sigma_D(\omega)$. By repeating our calculation under this assumption and comparing the two results, we have observed that their relative difference reaches values up to 115\% for $\mu=0\,$eV and $k=100\,\mu$m$^{-1}$ (being instead 27\% at $k=20\,\mu$m$^{-1}$), 7\% for $\mu=0.5\,$eV and $k=100\,\mu$m$^{-1}$ (1\% at $k=20\,\mu$m$^{-1}$). This shows that especially for low values of $\mu$ the role played by interband transitions (described by $\sigma_I(\omega)$ in eq. \eqref{Sigma}) is essential for a precise quantitative calculation. Going back to the analysis of graphene curves, we see that they allow to readily explain the behavior of the EM-LDOS shown in Fig. \ref{FigEM-LDOS}(b). First of all, the dispersion relation of plasmons on graphene does not have a horizontal asymptote in the $(\omega,k)$ plane \cite{NagaoPRL01} so that we do not expect a peak in the EM-LDOS. Yet, the EM-LDOS displays broad peaks which move when varying $\mu$. To explain their origin, we first observe that the integral describing the TM contribution to the electric density of states (eq. \eqref{EM-LDOSe}) is dominated by a factor $\kappa^2\exp(-2\kappa \omega z/c)$ for large $\kappa$ as $p$ is equivalent to $i\kappa$. This clearly shows that there is a distance-dependent cutoff wavevector $k_c=d^{-1}$ when observing at distance $d$. It is seen that the density of states has a peak value when $\kappa^2\exp(-2\kappa \omega d/c)$ is maximum, i.e. for $\kappa \omega/c=1/d$. This near-field filtering yields a \textit{$d$-dependent} value of the most represented wavevector $\kappa$. We can associate a frequency $\omega$ to this peak $\kappa$ value using the dispersion relation. In summary, the frequency peak in the EM-LDOS can be roughly estimated as the frequency associated to the largest participating wavevector, as confirmed by comparing Figs. \ref{FigEM-LDOS} and \ref{FigDR}.

We now finally discuss the case of graphene on SiC. The analysis of Fig. \ref{FigDR} shows the appearance of an anticrossing, proving that we are in presence of a strong coupling between SiC phonon-polariton and graphene plasmon as already shown \cite{KochPRB10,LiuPRB10}. The interplay between the optical properties of the two materials gives rise to two separate branches of dispersion relation, whose properties depend on the chemical potential. The most important effect due to the SiC-graphene coupling is the disappearance of the horizontal asymptote at $\omega=\omega_\text{spp}$. More specifically, for larger values of $\mu$ the high-frequency branch of the dispersion relation moves away from $\omega=\omega_\text{spp}$ for large values of $k$ and has no longer a flat asymptote. Conversely, the lower branch increases and remains bounded by a horizontal asymptote $\omega=\omega_T$. For $\mu=0.2\,$eV and $\mu=0.5\,$eV, the branch drifts toward higher values of frequencies, explaining the shift of the peak observed in Fig. \ref{FigEM-LDOS}. It is important to notice that the intersection between $k=d^{-1}=20\,\mu$m$^{-1}$ and the dispersion curve provides a rough estimate of the position of the modified peaks. Once again, the position of the EM-LDOS peak can be predicted from the dispersion relation, and the fact that this branch stays always below the one of suspended graphene corresponds to the fact that the peak in the EM-LDOS is always at lower frequencies for graphene-covered SiC with respect to suspended graphene.

\begin{figure}[h!]
\scalebox{0.0505}{\includegraphics{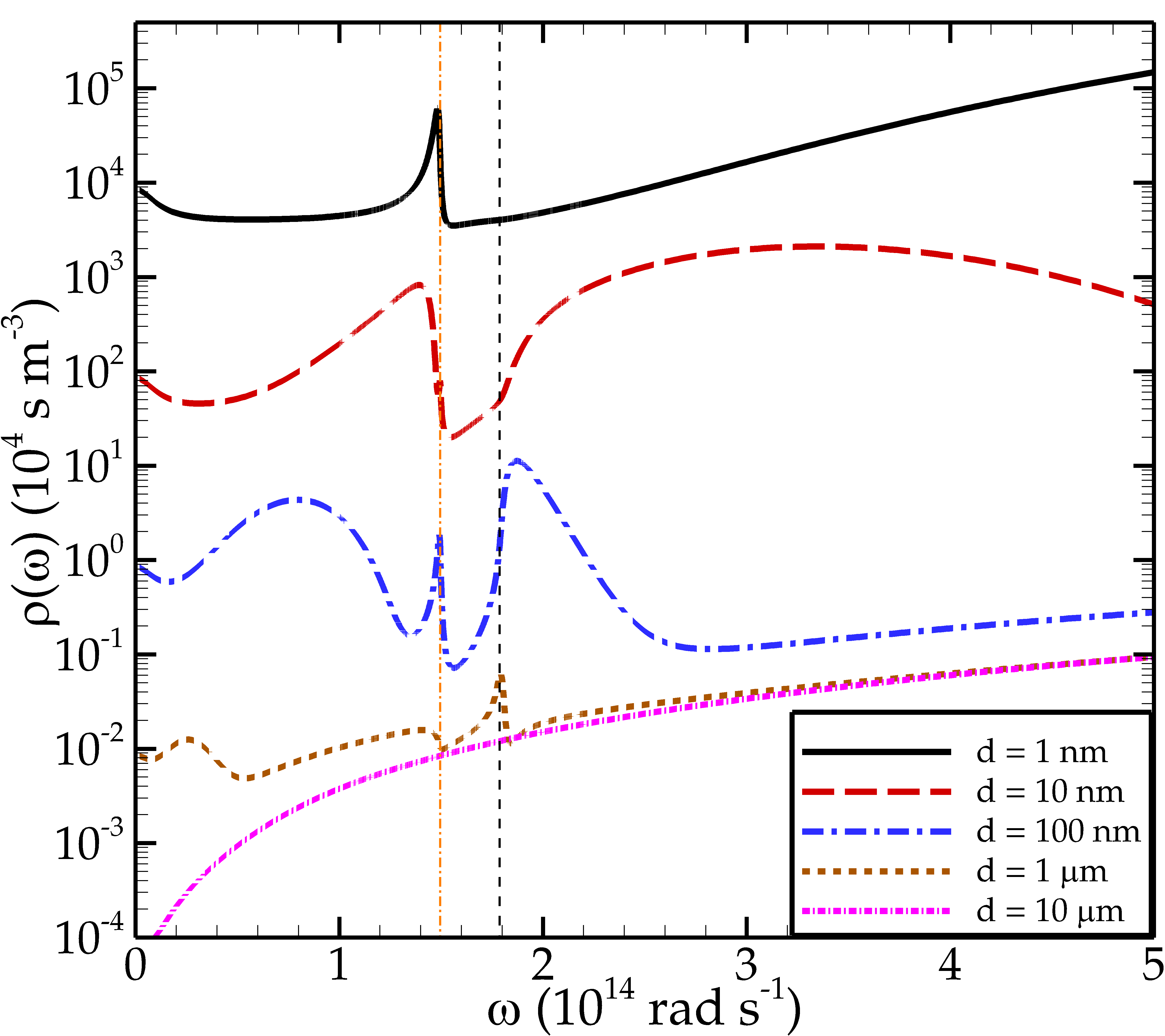}}
\caption{EM-LDOS for a SiC-graphene sample ($\mu=0.5\,$eV) for different values of the sample-tip distance $d$. The orange dot-dashed vertical line corresponds to $\omega=\omega_T$, whereas the dashed black vertical line is $\omega=\omega_\text{spp}$.}
\label{Figz}\end{figure}

We have highlighted that the position of the frequency peaks in the EM-LDOS can be predicted by the behavior of dispersion curves at the intersection points between $k=d^{-1}$ and the different branches of the coupled modes. Hence, this analysis predicts \textit{$d$-dependent} spectra of the EM-LDOS. We display in Fig. \ref{Figz} the EM-LDOS for a SiC-graphene system with $\mu=0.5\,$eV for several distances $d$ ranging from 1\,nm to $10\,\mu$m. It is seen that the spectra varies dramatically with distance. For SiC, the peak amplitude of the EM-LDOS varies but its position remains fixed. Here, owing to strong coupling between surface phonon polaritons and graphene plasmons, we observe peaks at frequencies which depend on the distance to the surface and on the chemical potential. This is a clear consequence of the interplay between strong coupling and near-field effects. To summarize, at a given distance $d$, we expect a dominant contribution to the EM-LDOS of the mode with $k=d^{-1}$. To this value correspond two frequency peaks given by Fig. \ref{FigDR}.  For instance, for the smallest considered distance $d=1\,$nm in Fig. \ref{FigEM-LDOS}, we have a cutoff $k_c\simeq 1000\,\mu$m$^{-1}$,  the quasi-horizontal shape of the lower branch of coupled modes (see Fig. \ref{FigDR}(c)), gives rise to a pronounced peak in proximity of $\omega=\omega_T$. In Fig. \ref{Figz}, we see that the EM-LDOS spectrum displays two peaks due to strong coupling for  $d>10\,$nm. On the contrary, beyond $1\,\mu$m the contribution of surface phonon polaritons and sheet plasmons to the EM-LDOS is reduced by five orders of magnitude so that the spectral peaks disappear. We conclude by emphasizing that the distance dependence of the spectrum has to be accounted for when designing experimental measurements of the EM-LDOS.

\section{Conclusions}\label{SecConcl}

We have calculated the electromagnetic local density of states in proximity of a graphene-covered SiC surface, and compared this result to the configurations of SiC and suspended graphene. This comparison has proved the appearance of new resonances and a strong dependence of the position of the new peaks both on the chemical potential of graphene and on the distance from the surface. We have also shown that the presence of the graphene sheet significantly broadens the EM-LDOS spectrum of the surface in the near field. All these features have been explained in terms of dispersion relations of surface modes, showing the occurrence of strong coupling between surface phonon polaritons and sheet plasmons of graphene. Hence, graphene can be used to tune the near-field optical behavior of SiC. This property paves the way to active control of the local density of states with possible applications to controlling quantum emitters lifetime or heat transfer at the nanoscale.

\end{document}